\documentclass[aps,pra,twocolumn, showpacs]{revtex4-1}

\usepackage{amsmath,amssymb}
\usepackage{graphicx}
\usepackage{epsfig}
\usepackage{multirow}
\usepackage{color}

\newcommand{\element}[3]{\langle #1|#2|#3\rangle}

\newcommand{\ket}[1]{|#1\rangle}
\newcommand{\ThreeJ}[6]{\left(\begin{array}{ccc}
#1&#2&#3\\
#4&#5&#6\\
\end{array}\right)}
\newcommand{\SixJ}[6]{\left\{\begin{array}{ccc}
#1&#2&#3\\
#4&#5&#6\\
\end{array}\right\}}

\newcommand{\sigv}{\hat{\sigma}_{\nu}}

\newcommand{\z}{\phantom{0}}
\newcommand{\vect}[1]{\vec{\boldsymbol{#1}}}
\newcommand{\bhat}[1]{\hat{\boldsymbol{#1}}}

\begin{document}

\input{epsf}

\title{Long-range interactions between rubidium and potassium Rydberg atoms}
\author{Nolan Samboy}

\affiliation{Department of Physical and Biological Sciences,\\ Western New England University,\\
            1215 Wilbraham Road, Springfield, MA 01119}

\date{\today}
\begin{abstract}
We investigate the long-range, two-body interactions between rubidium and potassium atoms in highly 
excited $(n=70)$ Rydberg states. After establishing properly symmetrized asymptotic basis states, 
we diagonalize an interaction Hamiltonian consisting of the standard Coulombic potential expansion 
and atomic fine structure to calculate
electronic potential energy curves. We find that when both atoms are excited to either the $70s$ state
or the $70p$ state, both the $\Omega=0+$ symmetry interactions and the $\Omega=0-$ symmetry interactions
demonstrate a deep potential well capable of supporting many bound levels; 
the size of the corresponding dimer states are on the 
order of 2.25~$\mu$m. We establish $n$-scaling relations for the equilibrium separation $R_e$ and the
dissociation energy $D_e$ and find these relations to be consistent with similar calculations
involving the homonuclear interactions between rubidium and cesium.
We discuss the specific effects of $\ell$-mixing and the exact composition of
the calculated potential well \textit{via} the expansion coefficients of the asymptotic 
basis states. Finally, we apply a Landau-Zener treatment to show that the dimer states are stable
with respect to predissociation.
\end{abstract}


\maketitle
\section{Introduction}
\label{sec:intro}
With the advent of laser cooling and atomic trapping, the investigation of Rydberg atoms
experienced a renaissance in the late 20$^{\rm th}$ century, 
which has led to many experimental discoveries and 
theoretical predictions. Exaggerated properties (long lifetimes, 
large cross sections, very large polarizabilities, etc.)~\cite{Gallagher} makes the 
Rydberg atom especially responsive
to external electric and magnetic fields, as well as to other Rydberg atoms.

Under ultracold conditions, the dipole-dipole
interactions between two Rydberg atoms is not masked by thermal motion, and so interactions
can occur at very long-range~\cite{Anderson,Mourachko}. These interactions 
have manifested in a variety of results including molecular resonance
excitation spectra~\cite{farooqi03,overstreet07}, ``exotic'' molecules (\textit{trilobite} 
states~\cite{trilobites,pfau} and \textit{macrodimer} 
states~\cite{macro-old,Samboy,Samboy-JpB,shaffer-NPHYS}), and the excitation-blockade
effect~\cite{lukin01}. All of these works uniquely illustrate the potential for applications
in quantum information processes (see~\cite{Saffman-RMP}, and more
recently~\cite{Marcassa2014}, for excellent comprehensive reviews of Rydberg physics research).

Within the last few years, the focus of study involving Rydberg systems has moved toward few-body 
interactions. For example, there have been proposals 
for the long-range interactions between one Rydberg atom and multiple ground
state atoms~\cite{Sadeghpour11,Sadeghpour10,Rost-poly,Jovica-trimers}, 
as well as for bound states between three Rydberg 
atoms~\cite{Samboy-trimer,jaksch-trimer13,jaksch-trimer14}. Currently,  
Rydberg states involving alkaline-earth elements are also being 
investigated~\cite{Vaillant12,Killian13,Killian16}, with the goal of forming Rydberg-Rydberg
pairs at large interatomic separations. The inner valence electrons in each atom of 
such a dimer would offer a new approach to probe and manipulate Rydberg systems.

The works mentioned here have all been 
with regard to homonuclear interactions; to the author's knowledge, Rydberg 
interactions between multiple species has not yet been considered. 
Fairly recently, photoassociation between different alkali species in the ground state 
($^{39}$K$^{85}$Rb) was achieved~\cite{StwalleyKRb1,StwalleyKRb2}. In principle,
such techniques could be applied to Rydberg states of these atoms to probe resonance features;
this paper aims to assist in such an effort. 

We discuss an approach 
for calculating long-range Rydberg interactions between two different alkali atoms;
although we
specifically discuss calculations involving rubidium and potassium, the theory can be applied to any
heteronuclear alkali pairing. Such results are relevant to the continuing work in ultracold
physics and chemistry, specifically with regard to the ``exotic'' Rydberg dimer states.
Please note: Except where otherwise indicated, atomic units are used throughout.
\section{Long-range Interactions}
\label{sec:longrange}
Neutral, alkali Rydberg atoms are convenient to explore because they are well-treated
using the semi-classical Bohr model 
(with the quantum defect correction)~\cite{Gallagher}.
In addition, the neutrality of the atoms ensures minimal interactions with the environment
while in the ground state~\cite{Calarco00,Vollbrecht04}, and 
the translational motion of the nuclei can be neglected
at ultracold temperatures~\cite{Mourachko, Gallagher-Frozen}.

When the distance between the two interacting Rydberg atoms is greater than the 
Le-Roy radius~\cite{LeRoy}:
\begin{equation}
\label{eq:LeRoy}
R_{LR}=2\,[\element{n_1\ell_1}{r^2}{n_1\ell_1}^{1/2}+
\element{n_2\ell_2}{r^2}{n_2\ell_2}^{1/2}]\,\,,
\end{equation}
the interactions are considered ``long-range'' and there is no overlap
of the two electron clouds. The potential energy of the interaction is then described by
that of two, well-separated charge distributions. For the case of Rydberg atoms, each
charge distribution is effectively a two particle system: a $+1$ nuclear core and a single,
highly excited valence electron.
\begin{figure}[h!]
	\centering
		\includegraphics[width=3.5in]{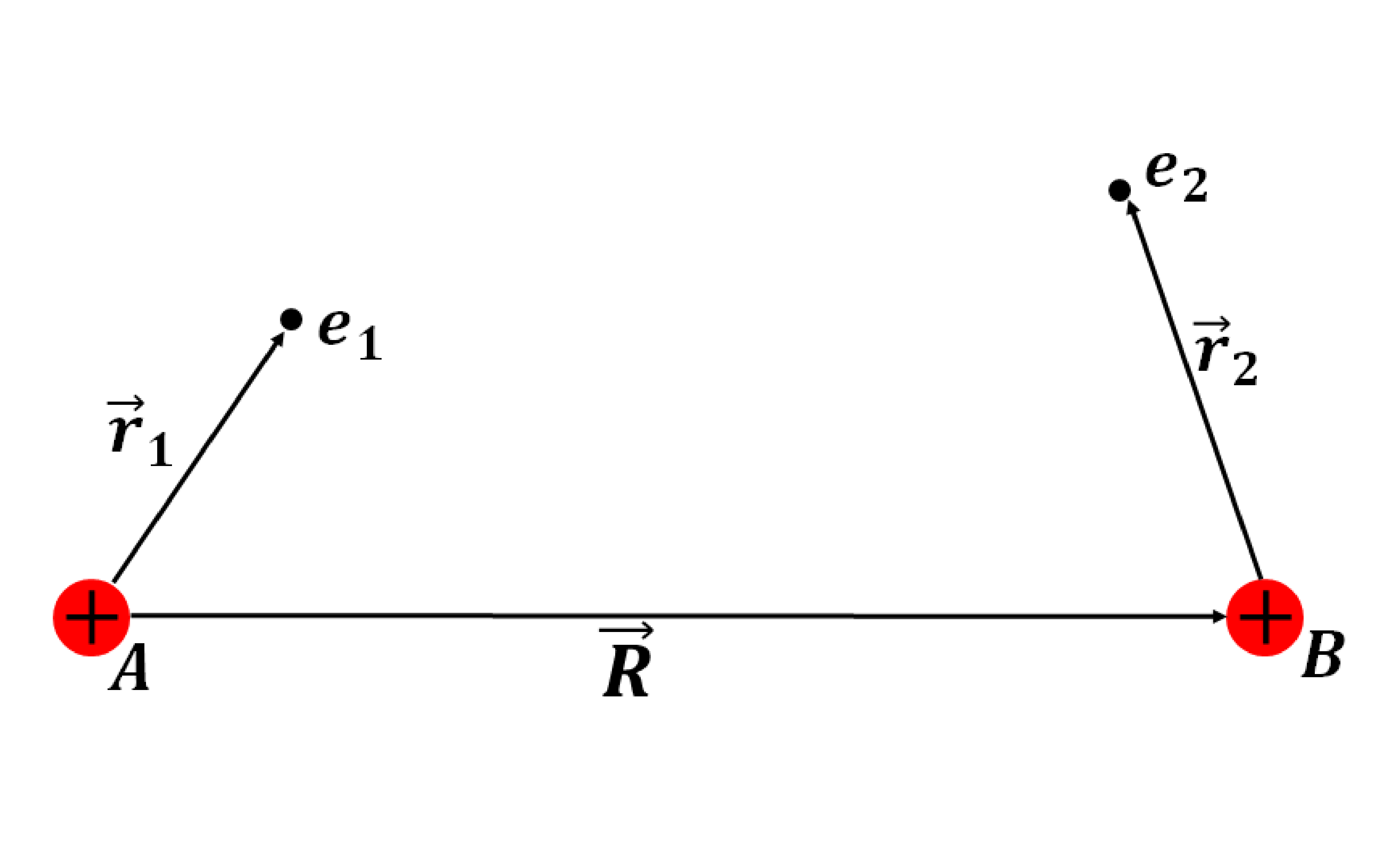}
	\caption{Two Rydberg atoms well separated from each other; each consists of a $+1$ nuclear core
	         and a single, highly excited valence electron $e_i$. Here, the nuclear distance 
					 $R$ is greater than the LeRoy radius (see text) and thus much larger
	         than either electron's distance $r_i$ from its respective nuclear core.}
	\label{fig:LongRange}
\end{figure}

Figure~\ref{fig:LongRange} schematically represents two such interacting Rydberg atoms:
electron 1 is a distance $r_1$ from core $A$, and electron 2 is a distance $r_2$ from core
$B$. In the long-range scenario, the nuclear distance $R$ is 
much larger than both $r_1$ and $r_2$. When the two nuclear cores are assumed to be fixed in space
(no kinetic energy), then
the interaction Hamiltonian is expressed in atomic units as:
\begin{equation}
\label{eq:Hamiltonian}
\hat{H} = \hat{H}_A+\hat{H}_B+\hat{V}_{\rm int}\,\,,
\end{equation}
where $\hat{H}_i$ contains the kinetic and potential energies of atom $i$ and $\hat{V}_{\rm int}$
is the Coulombic potential energy combinations of the two nuclei and the two electrons, given
in atomic units as:
\begin{equation}
\label{eq:Coulomb}
\hat{V}_{\rm int} = \dfrac{1}{R} - \dfrac{1}{|\vect{R}+\vect{r}_2|} - 
                    \dfrac{1}{|\vect{R}-\vect{r}_1|} + \dfrac{1}{|\vect{R}+\vect{r}_2-\vect{r}_1|}\,\,.
\end{equation}
\subsection{Basis States}
\label{subs:basis}
Given a non-relativistic Schr\"odinger equation, the fine-structure energy splitting is 
a result of the spin-orbit coupling between the total spin angular momentum of the
dimer $\vect{S}$  and the
total orbital angular momentum of the dimer $\vect{L}$. When nuclear rotation is neglected,
Hund's case (c) is the appropriate molecular
basis where the good quantum numbers are the total angular momentum of the dimer
$\vect{J} = \vect{L} + \vect{S}$ and its projection $\Omega$ along the internuclear axis.

We adopt a typical approach and assume the dimer wave function to be
a product of two atomic wave functions. Under the Born-Oppenheimer
approximation, each atomic wave function is solely
described by the quantum state of its valence electron 1 (2) about 
respective nucleus $A$ ($B$). In the coupled basis representation, each atom
possesses total atomic angular momentum $\vect{j}_i = \vect{\ell}_i+\vect{s}_i$,
where $\vect{\ell}_i$ is the orbital angular momentum of atom $i$ and $\vect{s}_i$ is the
spin angular momentum of atom $i$. The dimer wave functions are thus expressed as:
$\ket{1A}\equiv\ket{n_1,\ell_1,j_1,m_{j_1}}_{A}$ and 
$\ket{2B}\equiv\ket{n_{2},\ell_{2},j_{2},\Omega-m_{j_1}}_{B}$. Here $n_i$ is the principal
quantum number of atom $i$, $\ell_i$ is the orbital angular momentum quantum number of 
atom $i$, and $m_{j_i}$ is the projection of the total atomic angular momentum 
$\vect{j}_i$ of atom $i$ onto the internuclear axis (chosen in the $z$-direction for convenience). 

For the interactions considered here, the fine-structure energies are too
large for perturbation theory to be applicable. Thus, we directly
diagonalize the interaction Hamiltonian at successive values of $R$ to compute electronic
potential energy curves, as was done in~\cite{Jovica,Jovica-ndns}. Such an approach has
shown to be more successful at explaining experimental resonance features~\cite{farooqi03} 
because it more accurately describes the intricate mixing of the electrons'
angular momentum characters ($\ell$-mixing).

To facilitate faster computational times, we exploit molecular symmetries to construct
symmetrized molecular wave functions, as was done
in~\cite{Jovica,Jovica-ndns,Samboy,Samboy-JpB}. Although the heteronuclear dimer
does not possess  the inversion symmetry of its homonuclear counterpart, the 
total wave function does remain anti-symmetric with respect to electron exchange. 
Therefore,
as long as Equation~\eqref{eq:LeRoy} is satisfied, the properly symmetrized molecular
wave function is given by:
\begin{equation}
\label{eq:wf}
\ket{1A,2B;\Omega}\sim\frac{1}{\sqrt{2}}\left(\ket{1A}\ket{2B}-\ket{2B}\ket{1A}\right).
\end{equation}
For $\Omega=0$, reflection of the dimer through a plane containing the internuclear axis 
leads to wave functions that are either symmetric or anti-symmetric with respect to the
reflection operator $\sigv$. Furthermore, these wave functions have non-degenerate energy values
and must be uniquely defined. We distinguish between the two based on how $\sigv$ operates 
on~\eqref{eq:wf}:
\begin{equation}
\label{eq:sigv}
\ket{1A,2B;\Omega=0^{\pm}}=\left(\dfrac{1\pm\sigv}{\sqrt{2}}\right)\ket{1A,2B;\Omega=0}\,\,,
\end{equation}
where $\sigv$ behaves according to the following rules~\cite{Bernath,Brown}:
\begin{eqnarray}
\label{eq:refl-rules}
\sigv\ket{\Lambda}&=&(-1)^{\Lambda}\ket{-\Lambda}\\
\sigv\ket{S,M_S}&=&(-1)^{S-M_S}\ket{S,-M_S}\, .
\end{eqnarray}
\subsection{Basis sets}
In general, any basis set (defined by $\Omega$)
will consist of molecular states corresponding to those asymptotes
with significant coupling to both the Rydberg-Rydberg asymptotic level being considered 
and to other nearby states. We gauge the relative interaction strengths of local asymptotes based
on their contributions to the  
the $C_6 \sim \dfrac{(\element{\phi_1}{r}{\phi_2}\element{\phi_3}{r}{\phi_4})^2}{(E_1+E_3)-(E_2+E_4)}$
(dipole-dipole) and 
$C_5 \sim \element{\phi_1}{r^2}{\phi_2}\element{\phi_3}{r^2}{\phi_4}$ (quadrupole-quadrupole) 
coefficients of the molecular Rydberg state being considered. In these expressions, 
each $E_i$ is the asymptotic energy of atom $i$ in state
$\phi_i$ and $\element{\phi_i}{r^k}{\phi_j}$ is the radial matrix element between atoms $i$ and $j$.

As mentioned before, the $C_i$ coefficients (perturbation theory) are
not sufficient to properly describe the long-range Rydberg-Rydberg
interaction picture detailed in this work; however, such analysis does accurately assess 
which asymptotes provide strong coupling and which do not. 
For example, if we consider the $70^{(\rm K)}s+70^{(\rm Rb)}s$ Rydberg level
the largest contribution to the $C_6$ coefficient comes from the $69^{(\rm K)}p+70^{(\rm Rb)}p$
state ($8.89317\times 10^{14}$), while the largest contribution to the $C_5$ coefficient comes from 
the $67^{(\rm K)}d+69^{(\rm Rb)}d$ state ($3.8039\times 10^{15}$)
and the $68^{(\rm K)}d+70^{(\rm Rb)}d$ state ($2.4682\times 10^{15}$). 
To provide some contrast, the contribution of the $71^{(\rm K)}p+71^{(\rm Rb)}p$ state to the
$C_6$ coefficient is $2.51707\times 10^9$, while the contribution of the
$72^{(\rm K)}d+72^{(\rm Rb)}d$ state to the $C_5$ coefficient is $3.80343\times 10^{11}$. Since
the contributions of these states are 4 or 5 orders of magnitude smaller, they are not
included in the basis.

To construct a more complete basis set, we examine asymptotes in the vicinity 
$(\sim \pm 20$ GHz) of the molecular Rydberg level being considered and we
find in Table~\ref{tab:dipole} that the dipole strength between 
two atomic Rydberg states decays rapidly with the relative difference in their principal quantum numbers: 
$\Delta n \equiv |n_1 - n_2|$. Note: This table details specific results for transitions from
rubidium in the $70s$ state and from potassium in the $70s$ state, but similar behaviors are found
for transitions from any excited $n\ell$ Rydberg state.
Due to the sharp decline in the coupling strengths,
we only consider nearby asymptotic levels
whose two constituent atoms have $n_i$ values in the range $(n-3) \le n_i \le (n+3)$, where $n$
is the principal quantum number of the excited Rydberg state for each atom
($n=70$ for all tabulated results in this paper).
\begin{table}[h!]
   \centering
   \caption{Dipole matrix elements for atomic transitions from (left) Rb $70s$ and (right) K $70s$. 
            We highlight the two largest elements for each atomic species and note the rapid decrease 
            in coupling strength as $\Delta n=|n_1-n_2|$ increases.
            }
   \begin{tabular}{lcr|lcr}
   \vspace{1pt}\\
   \hline\hline
	Rubidium & & & Potassium & & \\
	\hline
   $\element{70s}{r}{74p_{3/2}}$ & $=$ & 94.064 & $\element{70s}{r}{74p_{3/2}}$ & $=$ & 87.093 \\
   $\element{70s}{r}{73p_{3/2}}$ & $=$ & -144.19 & $\element{70s}{r}{73p_{3/2}}$ & $=$ & -134.41 \\
   $\element{70s}{r}{72p_{3/2}}$ & $=$ & 258.08 & $\element{70s}{r}{72p_{3/2}}$ & $=$ & 243.23\\
   $\element{70s}{r}{71p_{3/2}}$ & $=$ & -639.57 & $\element{70s}{r}{71p_{3/2}}$ & $=$ & -616.30 \\
   $\element{70s}{r}{70p_{3/2}}$ & $=$ & \textbf{5081.6} & $\element{70s}{r}{70p_{3/2}}$ & $=$ & \textbf{5353.1} \\
   $\element{70s}{r}{69p_{3/2}}$ & $=$ & \textbf{4807.8} & $\element{70s}{r}{69p_{3/2}}$ & $=$ & \textbf{4810.7} \\
   $\element{70s}{r}{68p_{3/2}}$ & $=$ & -649.29 & $\element{70s}{r}{68p_{3/2}}$ & $=$ & -696.92 \\
   $\element{70s}{r}{67p_{3/2}}$ & $=$ & 262.07 & $\element{70s}{r}{67p_{3/2}}$ & $=$ & 285.92 \\
   $\element{70s}{r}{66p_{3/2}}$ & $=$ & -144.50 & $\element{70s}{r}{66p_{3/2}}$ & $=$ & -158.83 \\
   \hline\hline
   \end{tabular}
\label{tab:dipole}
\end{table}
Table~\ref{tab:K} lists the relevant molecular levels near the $70s+70s$ asymptote and the
$70p+70p$ asymptote; the properly symmetrized states corresponding to these molecular
levels comprise the appropriate basis sets.
\begin{table}[h]
\caption{Molecular asymptotes included in the basis sets for the long-range interactions between 
         rubidium and potassium: (left) each atom excited to the $70s$ Rydberg state, and (right) 
				 each atom excited to the $70p$ Rydberg state.
				}
\begin{tabular}{c|c}
\hline\hline
K$(70s)$-Rb$(70s)$ & K$(70p)$-Rb$(70p)$\\
\hline
$68^{(\rm K)}d+70^{(\rm Rb)}s$ & $67^{(\rm K)}f+71^{(\rm Rb)}p$\\
$69^{(\rm K)}s+69^{(\rm Rb)}d$ & $68^{(\rm K)}f+70^{(\rm Rb)}p$\\
$68^{(\rm K)}p+71^{(\rm Rb)}p$ & $68^{(\rm K)}d+69^{(\rm Rb)}d$\\
$69^{(\rm K)}p+70^{(\rm Rb)}p$ & $69^{(\rm K)}s+70^{(\rm Rb)}d$\\
$69^{(\rm K)}s+71^{(\rm Rb)}s$ & $67^{(\rm K)}d+72^{(\rm Rb)}s$\\
$70^{(\rm K)}s+70^{(\rm Rb)}s$ & $69^{(\rm K)}d+70^{(\rm Rb)}s$\\
$68^{(\rm K)}d+69^{(\rm Rb)}d$ & $68^{(\rm K)}d+71^{(\rm Rb)}s$\\
                               & $69^{(\rm K)}p+71^{(\rm Rb)}p$\\
                               & $70^{(\rm K)}p+70^{(\rm Rb)}p$\\
                               & $69^{(\rm K)}s+72^{(\rm Rb)}s$\\
                               & $70^{(\rm K)}s+71^{(\rm Rb)}s$\\
                               & $67^{(\rm K)}f+69^{(\rm Rb)}f$\\
                               & $68^{(\rm K)}f+68^{(\rm Rb)}f$\\
\hline\hline
\end{tabular}
\label{tab:K}
\end{table}

\subsection{Interaction Hamiltonian}
\label{subs:IntHam}
Under the Born-Oppenheimer approximation, diagonalization of the interaction Hamiltonian
(Eq.~\eqref{eq:Hamiltonian})
results in a set of electronic energies with regard to a fixed nuclear separation $R$. A complete set of 
electronic energy curves can be calculated by diagonalizing a unique Hamiltonian matrix at varying values of $R$.

A convenient approach for long-range Rydberg investigations is to express the Coulombic potential 
energy expression (Eq.~\eqref{eq:Coulomb}) as a multipole expansion in inverse powers of 
$R$~\cite{rose,Hirschfelder,Rushbrooke}; 
the expansion is further simplified if we assume $\vect{R}$ lies along a $z-$axis, common to both
Rydberg atoms~\cite{Fontana}:
\begin{align}
\label{eq:multipole}
\hat{V}_{\rm int}\equiv V_L(R)&= {\displaystyle\sum_{L = 0}^{\infty}}
                    (-1)^{L} \dfrac{ 4\pi }{R^{2L + 1}(2L+1)} r_1^{L}r_2^{L} \nonumber\\
								&\times	{\displaystyle\sum_{m = -L}^{L}}
									 B_{2L}^{L+m}\,Y_{L}^{m}(\bhat{r}_1)\,Y_{L}^{-m}(\bhat{r}_2)\,\,.
\end{align}
Here, $B_n^k \equiv \frac{n!}{k!(n-k)!}$ is the binomial coefficient, $Y_L^m(\bhat{r}_i)$ is
a spherical harmonic describing the angular position of electron $i$ with position $\vect{r}_i$ from
its nuclear center, $L$ labels the ($2^L$) multi-pole moment ($L=1$ for dipolar, $L=2$ for quadrupolar, etc.), 
and $R$ is the
internuclear distance. The advantage to such an expansion 
is that the expression can be truncated such that only meaningful terms are kept; this significantly
reduces computation time.

When the Hamiltonian is diagonalized, the expectation value of each term in the energy expansion is
proportional to $\dfrac{\langle r_1^L \rangle \langle r_2^L \rangle}{R^{2L+1}}$; for Rydberg atoms,
each radial element scales as $\langle r^L \rangle \sim n^{2L}$\cite{Gallagher}. Thus, dipole-dipole
interactions scale as $\sim \dfrac{n^4}{R^3}$, quadrupole-quadrupole interactions scale as 
$\sim \dfrac{n^8}{R^5}$ and so on. For the internuclear spacings considered here, $R\approx n^{5/2}$,
so the dipole-dipole coupling strength is $\sim n^{-7/2}$ and the quadrupole-quadrupole coupling strength
is $\sim n^{-9/2}$. 

Typically, dipole-dipole interactions dominate the long-range Rydberg-Rydberg interactions,
but it has been shown~\cite{Jovica-ndns,shaffer} that quadrupole-quadrupole couplings can also be
significant to the interaction picture. To date, octupole-octupole
interactions have not been shown to be relevant in long-range Rydberg interactions, 
so we do not consider them here. Based on the scaling relations shown above,
such a term would be a factor of $n^{-1}$
less than the quadrupole-quadrupole term and a factor of $n^{-2}$ less than the dipole-dipole
term. 

Because the molecular basis states are linear combinations of atomic states determined
through symmetry considerations (see Eq.~\eqref{eq:wf}), each matrix element in the interaction
Hamiltonian is actually
a combination of multiple interaction terms:
\begin{align}
\label{eq:VintExp}
\element{1A,2B;\Omega}{&\hat{H}}{3A,4B;\Omega} \sim \nonumber\\
           &\element{1A,2B}{\hat{H}}{3A,4B} \nonumber\\
          -&\element{1A,2B}{\hat{H}}{4A,3B} \\
          -&\element{2A,1B}{\hat{H}}{3A;4B} \nonumber\\ 
          +&\element{2A,1B}{\hat{H}}{4A;3B} \;.\nonumber
\end{align}
For the $\Omega=0$ case, equation~\eqref{eq:sigv} is also applied, resulting in additional 
terms. Since the normalization factor varies with state definitions and symmetry considerations,
it is not stated explicitly in~\eqref{eq:VintExp}. Normalization factors are included in calculations,
however.
In this notation, \\
$\ket{1A,2B} =\ket{n_1,\ell_1,j_1,m_{j_1}}_A\ket{n_2,\ell_2,j_2,\Omega-m_{j_1}}_B$, and so on.
\begin{figure}[h!]
\centering
   \includegraphics[width=3.5in]{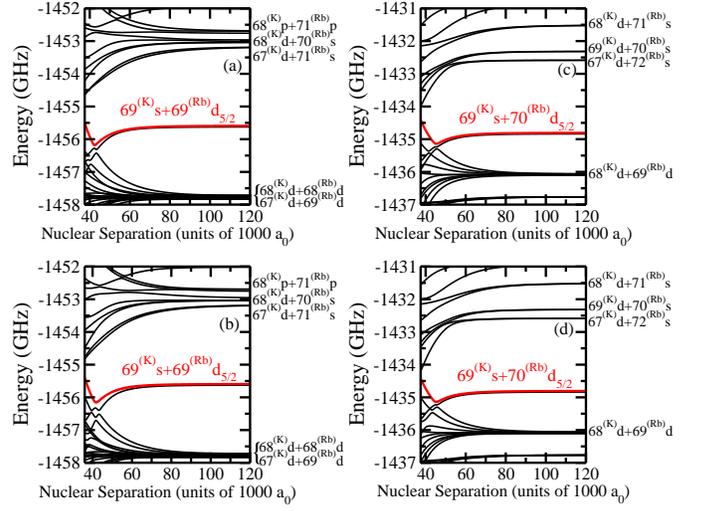}
\caption{(Color online) Long-range potential energy curves corresponding to the interactions between one 
         rubidium atom
         and one potassium atom excited to the same state. The curves
				 in panel (a) correspond to the $\Omega = 0^+$ symmetry with both atoms excited to the $70s$ state, 
				 the curves in panel (b) correspond 
				 to the $\Omega = 0^-$ symmetry with both atoms excited to the $70s$ state,
				 the curves in panel (c) correspond to the 
				 $\Omega = 0^+$ symmetry with both atoms excited to the $70p$ state and the curves in panel (d)
				 correspond to the $\Omega = 0^-$ symmetry with both atoms excited to the $70p$ state. We highlight the 
				 potential energy wells in red and label their respective asymptotic states.
         }
\label{fig:RbK}
\end{figure}

An analytical expression for any given term in the matrix element is found to be:
\begin{align}
\label{eq:element}
&\element{1A,2B}{V_L(R)}{3A,4B}=\nonumber\\
&(-1)^{L-1-\Omega+j_{\rm tot}}
\sqrt{\hat{\ell}_1\hat{\ell}_2\hat{\ell}_3\hat{\ell}_4\hat{j}_1\hat{j}_2\hat{j}_3\hat{j}_4}
\;\dfrac{\mathcal{R}_{13,A}^L\;\mathcal{R}_{24,B}^L}{R^{2L+1}}\nonumber\\
&\times\ThreeJ{\ell_1}{L}{\ell_3}{0}{0}{0}\ThreeJ{\ell_2}{L}{\ell_4}{0}{0}{0}\nonumber\\
&\times\SixJ{j_1}{L}{j_3}{\ell_3}{\frac{1}{2}}{\ell_1}
\SixJ{j_2}{L}{j_4}{\ell_4}{\frac{1}{2}}{\ell_2}\\
&\times{\displaystyle\sum_{m=-L}^{L}}B_{2L}^{L+m}\ThreeJ{j_1}{L}{j_3}{-m_{j_1}}{m}{m_{j_3}}\nonumber\\
&\times\ThreeJ{j_2}{L}{j_4}{-m_{j_2}}{-m}{m_{j_4}}\,\,,\nonumber
\end{align}
where $j_{\rm tot}=j_1+j_2+j_3+j_4$, $\hat{\ell}_i=2\ell_i+1$, $\hat{j}_i=2j_i+1$,
and $\mathcal{R}_{ij,Q}^L=\element{i}{r^L}{j}_Q$ is the radial matrix element of atom $Q$.
The $(\cdots )$ expressions represent Wigner-$3j$ symbols, and the $\{\cdots\}$ represent
Wigner-$6j$ symbols.

The diagonal elements, \textit{i.e.} $\ket{1A,2B}~=~\ket{3A,4B}$ are given by:
\begin{align}
\element{1A,2B;\Omega}{\hat{H}}{&1A,2B;\Omega}=\nonumber\\
&\element{1A,2B;\Omega}{V_L(R)}{1A,2B;\Omega} \nonumber\\
+ &\,\,E_{1A}+E_{2B}\,,
\label{eq:diag}
\end{align}
where $\element{1A,2B;\Omega}{V_L(R)}{1A,2B;\Omega}$ follows~\eqref{eq:VintExp} 
and~\eqref{eq:element}, and
each $E_{iQ}$ is the asymptotic energy of the atomic Rydberg state
$\ket{n_i,\ell_i,j_i,m_{j_i}}_Q$:
\begin{equation}
\label{eq:RydEn}
E_{iQ} = -\dfrac{1}{2(n_{iQ} -\delta_{\ell_i Q})}\,\,.
\end{equation}
In this expression, $n_{iQ}$ is the principal quantum number of atom $Q$ and $\delta_{\ell_i Q}$ 
is the quantum defect for atom $Q$ (values given in~\cite{Gallagher,Li,nfqd}).
\section{Interaction Curves}
After investigating all possible $\Omega$-symmetries for the $70s+70s$ and $70p+70p$ excitations of 
rubidium and potassium, we found that in both cases the $\Omega = 0^+$ and the $\Omega = 0^-$ symmetries resulted
in potential wells capable of supporting bound states. In Figure~\ref{fig:RbK}, we plot the interaction
energies for these four cases against the Bohr radius ($a_0$) and highlight the resulting
potential wells in red; we also label the wells' corresponding asymptotic energy levels.

\begin{figure}[h!]
	\centering
		\includegraphics[width=3.5in]{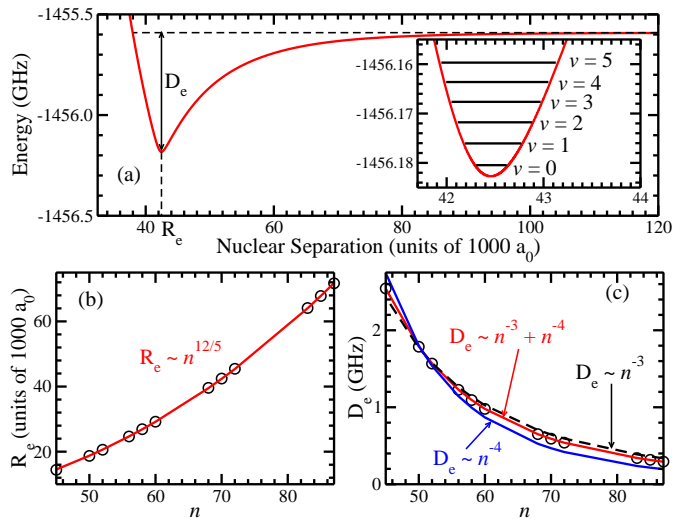}
	\caption{(Color online) (a) Long-range potential energy curves corresponding to the 
	         $\Omega=0^+$ symmetry
	         interactions of one potassium Rydberg
	         atom and one rubidium Rydberg atom; both atoms have been excited to the 70$s$ state. 
					 We note the existence
					 of a $\sim$ 600 MHz deep well (associated with the $69^{(\rm K)}s + 69^{(\rm Rb)}d_{5/2}$ 
					 asymptotic level),
					 capable of supporting many bound states. We explicitly label the equilibrium separation $R_e$
					 and the dissociation energy $D_e$.
					 Inset: Zoom of the deepest part of the well, 
					 showing the first few bound vibrational levels; corresponding energies and classical 
					 turning points are given in Table~\ref{tab:bound}. (b) Scaling relations for the equilibrium
					 separation $R_e$ vs. principal quantum number $n$: $R_e \sim n^{12/5}$. 
					 (c) Scaling relations for the 
					 dissociation energy $D_e$ vs. principal quantum number $n$. Three different results (see text) 
					 are shown: $D_e\sim n^{-4}$ (blue) gives poor agreement, $D_e\sim n^{-3}$ (black-dashed) gives
					 good agreement, but $D_e\sim n^{-3}+n^{-4}$ (red) gives the best agreement.
	         }
	\label{fig:KRbwell}
\end{figure}

For all four of these wells, we explored scaling relations, composition, and stability. We provide
a visual example in Figure~\ref{fig:KRbwell}(a), where the potential well corresponding to 
the $\Omega=0^+$ symmetry for both atoms excited to the $70s$ state is isolated.
This well is $\sim 600$ MHz deep and supports $\sim 630$ bound vibrational states. In
Table~\ref{tab:bound}, we present the first few bound state vibrational energies 
(as measured from the bottom of the well) for all four wells, as well as the classical turning points, 
indicating the large size of these dimer states.
Given that the equilibrium separation $R_e$ for all four wells is between 41,000~$a_0$ and
46,000~$a_0$ $(\sim 2.25$~$\mu$m),
these bound states are very extended, consistent with the \textit{macrodimer} classification.

The inset of Figure~\ref{fig:KRbwell}(a) shows that the deepest part of the wells can be well-modeled as a 
harmonic potential; the first few bound levels for each well are consistently spaced (see Table~\ref{tab:bound}).
The bound energies and corresponding wave functions were calculated using the 
mapped Fourier Grid Method~\cite{grid} for each potential well individually.
\subsection{Scaling Relations}
The dissociation energy $D_e$ and the equilibrium separation $R_e$ for the potential wells were
were calculated for various 
values of the principal quantum number $n$. In panels (b) and (c) of Figure~\ref{fig:KRbwell},
we show the results for the $\Omega = 0^+$ symmetry with both atoms excited to the $70s$ state. 
In panel (b), we see that the equilibrium
separation scales as $n^{12/5}$, and in panel (c), we present different ``best-fit'' curves corresponding to 
different values of $n$-scaling for the dissociation energy. 
For pure dipole-dipole coupling, one would expect the dissociation energy to 
scale as the energy difference between energy levels ($n^{-3}$ for Rydberg atoms~\cite{Gallagher}).
We see that this result gives pretty good agreement, but the results suggest that $D_e$ 
actually scales
as $\sim n^{-4}+n^{-3}$. Although it gives poor agreement, we also include the curve of $n^{-4}$ 
for completeness.
In Table~\ref{tab:scaling}, we present the scaling results for all four of the potential wells identified in 
Figure~\ref{fig:RbK}. We note that the scaling results obtained here for 
$R_e$ and $D_e$ are consistent with the results for homonuclear macrodimers, presented in~\cite{Samboy-JpB}. 
A thorough derivation for these scaling relations was also performed in that work; 
we do not republish them here.
\begin{table}[h]
   \caption{$n$-scaling relations for the equilibrium separation $R_e$ and the dissociation energy $D_e$ for the
	          four potential wells discussed in this work (see Figure~\ref{fig:RbK}). 
						Both results are consistent with previous work regarding
						homonuclear macrodimers (see text).
            }
   \begin{tabular}{cclr}
   \hline\hline
 Threshold energy (well) & Symmetry & $R_e$ scaling & $D_e$ scaling\\
   \hline
 $69^{(\rm K)}s+69^{(\rm Rb)}d_{5/2}$ & $\Omega = 0^+$ & 
                                        $n^{2.4}$ ($n^{12/5}$) & $n^{-3}+n^{-4}$\\
$69^{(\rm K)}s+69^{(\rm Rb)}d_{5/2}$ & $\Omega = 0^-$ &
                                         $n^{2.4}$ ($n^{12/5}$) & $n^{-3}+n^{-4}$\\
 $69^{(\rm K)}s+70^{(\rm Rb)}d_{5/2}$ & $\Omega = 0^+$ &
                                         $n^{2.3}$ & $n^{-3}+n^{-4}$\\
$69^{(\rm K)}s+70^{(\rm Rb)}d_{5/2}$ & $\Omega = 0^-$ & 
                                         $n^{2.3}$ & $n^{-3}+n^{-4}$\\
   \hline\hline
   \end{tabular}
\label{tab:scaling}
\end{table}   
\subsection{Well composition and stability}
\begin{figure}[t]
	\centering
		\includegraphics[width=3.5in]{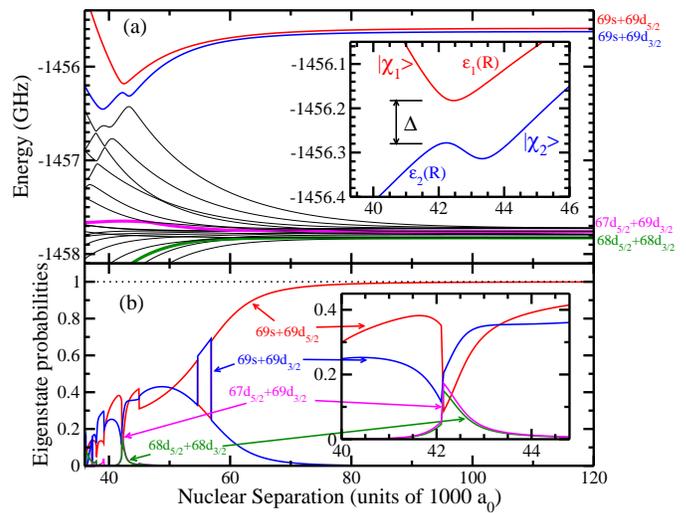}
	\caption{(Color Online) (a) $\Omega = 0^+$ potential energy curves describing the 
	          interactions between one potassium
	          Rydberg atom and one rubidium Rydberg atom: both atoms have been excited to the $70s$
						state. 
						We highlight those curves corresponding to the electronic states that 
						contribute the most to the formation
						of the well (see text). Inset: Zoom of the region near the avoided crossing, indicating
						the relevant features of the Landau-Zener
						treatment: $\Delta$ is the energy gap between the adiabatic
						energies $\epsilon_1$ and $\epsilon_2$ at closest approach; these adiabatic energies
						correspond to the two adiabatic states $\ket{\chi_1}$ and $\ket{\chi_2}$ at the avoided crossing.
						(b) Probabilities 
						$\left|c_j(R)\right|^2$ of the electronic basis states $\ket{j}$ most responsible for the formation
						of the well (see text). After $\sim$ 60,000~$a_0$, the $\ell$-mixing is negligible: the probability
						coefficient corresponding to $69^{\rm (K)}s+^{(\rm Rb)}70d_{5/2}$ 
						approaches one (see dashed line) and all other probability coefficients decay to zero. 
						Inset: Zoom of the region near the avoided crossing.
						All panels use the same labeling/color scheme.	
	         }
	\label{fig:EigVex}
\end{figure}

Due to the electronic $\ell$-mixing, each potential energy curve $U_{\lambda}(R)$ 
is described by an electronic wave function $\ket{\chi_{\lambda}(R)}$, which itself is 
a superposition of the asymptotic molecular wave functions~\eqref{eq:wf}: 
\begin{equation}
\ket{\chi_{\lambda}(R)} = {\displaystyle\sum_j c_j^{(\lambda)}(R)}\ket{j}\,\,.
\end{equation}
The exact amount of mixing varies with $R$, and is completely described by the $c_j^{(\lambda)}(R)$ 
coefficients: the eigenvectors after diagonalization. The $\ket{j}$ are the corresponding
symmetrized basis states $\ket{1A,2B;\Omega}$, defined earlier~\eqref{eq:wf}. 
As an example of the $R$-dependence of the mixing, in Figure~\ref{fig:EigVex}(b)
we illustrate the composition of the potential well highlighted in Figures~\ref{fig:KRbwell}(a)
and~\ref{fig:EigVex}(a). This potential curve corresponds to the $69^{(\rm K)}s + 69^{(\rm Rb)}d_{5/2}$
asymptote.

As would be expected, this well is mainly composed of the $69^{(\rm K)}s + 69^{(\rm Rb)}d_{5/2}$ 
state. However, in the region of the actual well, we also see significant contributions from the
$69^{(\rm K)}s + 69^{(\rm Rb)}d_{3/2}$ directly below the well, and even from 
some deeper $nd+n'd$ states. In Figure~\ref{fig:EigVex}(a), we highlight and label
the four states most relevant to $\ell$-mixing;
the corresponding probabilities $\left|c_j(R)\right|^2$ of these
states are plotted against $R$ in Figure~\ref{fig:EigVex}(b) to explicitly describe the 
$R$-dependence. We note that the effects from $\ell$-mixing cease when the nuclear separation is about 
60,000~$a_0$. This is obvious by the probability coefficient of the
$69^{(\rm K)}s + 69^{(\rm Rb)}d_{5/2}$ state approaching one, while all other coefficients approach zero.
Also of note is the switch in probabilities of the $69^{(\rm K)}s + 69^{(\rm Rb)}d_{5/2}$ and
$69^{(\rm K)}s + 69^{(\rm Rb)}d_{3/2}$ states between $\sim$ 55,000 $a_0$ and $\sim$ 57,000 $a_0$.
Although perhaps not visually obvious in the Figure~\ref{fig:EigVex}(a) panel, these two curves do appear
to experience an avoided crossing in this general vicinity: one can observe a slight deviation in
the line shapes of these two curves as they meet. Such a crossing would be consistent with the behavior
of the probabilities.

For potential energy wells formed \textit{via} some avoided crossing between curves, predissociation
of bound energy levels can be a concern. In general, avoided crossings can lead to predissociation
if the (meta-)stable state has strong coupling to the unstable state below it. From the eigenvector
plot in Figure~\ref{fig:EigVex}(b), we see that there is significant coupling between the potential well
and the curves that lie below it. Therefore, the long-term stability of these wells could be compromised.
\begin{table*}[t!]
   \caption{Energies of the deepest bound levels (measured from the bottom of the well),
            classical turning points ($R_1$, $R_2$), and dimer lifetimes
						$\tau$ (see text) for the four potential wells identified in 
						Figure~\ref{fig:RbK}.
            }
   \begin{tabular}{cccccccc}
   \hline\hline
  Excited Rydberg pair & \phantom{sp}Threshold energy (well)\phantom{sp} & Symmetry
	        & $v$ & Energy (MHz) & $R_1$ (a.u.) & $R_2$ (a.u.) & $\tau$ (s)\\
   \hline
$70^{(\rm K)}s+70^{(\rm Rb)}s$ & $69^{(\rm K)}s+69^{(\rm Rb)}d_{5/2}$ & $\Omega = 0^+$ & 0 
     & \z\z 2.2240 & 42 175 & 42 783  &  $\infty$ \\
 & & &  1 & \z\z 6.6490 & 42 175  & 42 784 & $\infty$ \\
 & & &  2 & \z 10.9194 & 42 095 & 42 891 & $\infty$\\
 & & &  3 & \z 15.0636 & 42 031 & 42 985 & $\infty$\\
 & & &  4 & \z 19.0879 & 41 976 & 43 070 & $\infty$\\
 & & &  5 & \z 23.0177 & 41 926 & 43 149 &  $\infty$\\
&	& &  $\vdots$ & $\vdots$ & $\vdots$ & $\vdots$ & $\vdots$ \\
&	& & 627 & 591.9279 & 37 967 & 226 597 & $\infty$\\
\hline
$70^{(\rm K)}s+70^{(\rm Rb)}s$& $ 69^{(\rm K)}s+69^{(\rm Rb)}d_{5/2}$ & $\Omega = 0^-$ & 0 
     & \z\z 1.7120 & 41 591 & 41 983 & $\infty$\\
 & & &  1 & \z\z 5.1375 & 41 456 & 42 145 & $\infty$ \\
 & & &  2 & \z\z 8.5053 & 41 366 & 42 260 & $\infty$\\
 & & &  3 & \z 11.8269 & 41 292 & 42 360 & $\infty$\\
 & & &  4 & \z 15.0967 & 41 232 & 42 447 &  $\infty$\\
 & & &  5 & \z 18.3146 & 41 177 & 42 527 &  $\infty$\\
&	& &  $\vdots$ & $\vdots$ & $\vdots$ & $\vdots$ & $\vdots$ \\
&	& & 608 & 564.7000 & 37 067 & 157 474 & $\infty$\\	
\hline
$70^{(\rm K)}p+70^{(\rm Rb)}p$ & $69^{(\rm K)}s+70^{(\rm Rb)}d_{5/2}$ & $\Omega = 0^+$ & 0 
     & \z\z 1.3422 & 45 088 & 45 535 & $\infty$\\
 & & &  1 & \z\z 3.9897 & 44 937 & 45 718 & $\infty$ \\
 & & &  2 & \z\z 6.5859 & 44 834 & 45 854 & $\infty$\\
 & & &  3 & \z\z 9.1309 & 44 750 & 45 970 & $\infty$\\
 & & &  4 & \z 11.6247 & 44 681 & 46 073 &  $\infty$\\
 & & &  5 & \z 14.0752 & 44 620 & 46 169 &  $\infty$\\
&	& &  $\vdots$ & $\vdots$ & $\vdots$ & $\vdots$ & $\vdots$ \\
&	& & 470 & 335.6738 & 40 716 & 181 441 & $\infty$\\
\hline
$70^{(\rm K)}p+70^{(\rm Rb)}p$ & $69^{(\rm K)}s+70^{(\rm Rb)}d_{5/2}$ & $\Omega = 0^-$ & 0 
     & \z\z 1.3682 & 45 072 & 45 512 & $\infty$\\
&  & &  1 & \z\z 4.0782 & 44 921 & 45 696  & $\infty$ \\
&  & &  2 & \z\z 6.7330 & 44 821 & 45 832 & $\infty$\\
&  & &  3 & \z\z 9.3245 & 44 739 & 45 947 & $\infty$\\
&  & &  4 & \z 11.8687 & 44 670 & 46 050 &  $\infty$\\
&  & &  5 & \z 14.3575 & 44 608 & 46 147 &  $\infty$\\
&	& &  $\vdots$ & $\vdots$ & $\vdots$ & $\vdots$ & $\vdots$ \\
&	& & 473 & 337.7708 & 40 746 & 158 929 & $\infty$\\	
   \hline\hline
   \end{tabular}
\label{tab:bound}
\end{table*}   

For curves that are less intricate, a simple Landau-Zener~\cite{Landau, Zener} treatment would be
desirable. However, such an approach depends on detailed knowledge of the \textit{diabatic} crossing
behavior, including which diabatic states actually correspond to the crossing potential energy curves; 
for the complicated curve mixings that we demonstrate, defining these diabatic states becomes difficult.
Instead, we adopt the approach taken by Clark~\cite{Clark}, in which the parameters defining the Landau-Zener
probability are obtained from the adiabatic $P$-matrix coupling. Specifically, the $P$-matrix is defined through
the off-diagonal derivative of the interaction potential between the crossing states:
\begin{equation}
P_{12}(R) = \dfrac{\element{\chi_1}{(\partial/\partial R)V(R)}{\chi_2}}
                  {\left(\varepsilon_1(R) - \varepsilon_2(R)\right)}\,\,.
\end{equation}
Here, $\varepsilon_i(R)$ is the energy value of the adiabatic potential curve described by state 
$\ket{\chi_i}$, and  $V(R)$ is defined by equation~\eqref{eq:multipole}. As the nuclear separation is varied, the
value of the $P$-matrix peaks when the adiabatic curves are at their closest approach
(corresponding to $\varepsilon_1(R)-\varepsilon_2(R)$ being a minimum). Clark showed that
by fitting the $P$-matrix to a Lorentzian function, the probability $P_{LZ}$ to make a non-adiabatic transition 
from the electronic state $\ket{\chi_1}$ to the
electronic state $\ket{\chi_2}$ is given by:
\begin{equation}
P_{LZ} = \exp{\left(-\dfrac{2\pi}{v}\dfrac{\Delta}{8\,P_{\rm max}}\right)}\,\,,
\label{eq:LZ}
\end{equation}
where $v$ is the relative velocity of the two nuclei determined from the bound levels of the molecule,
$\Delta$ is the energy gap between the two adiabatic curves at closest approach, 
and $P_{\rm max}$ is the peak value of the $P$-matrix. 
The inset of Figure~\ref{fig:EigVex}(a)
shows a close-up of the avoided crossing between the 
$\ket{\chi_1}\equiv 69^{(\rm K)}s + 69^{(\rm Rb)}d_{5/2}$  
and $\ket{\chi_2}\equiv 69^{(\rm K)}s + 69^{(\rm Rb)}d_{3/2}$ 
electronic states; the energy gap $\Delta$ is also indicated.\\
\\
Since $P_{LZ}$ represents the likelihood of transitioning from $\ket{\chi_1}$ to $\ket{\chi_2}$ (and
thus predissociating into two free atoms), then $1-P_{LZ}$ is the probability that the macrodimer will
remain in $\ket{\chi_1}$ and \textit{not} predissociate. We match this probability to an exponential
decay over the time $t$ for a full oscillation inside the well: $1-P_{LZ} = e^{-t/\tau}$ and find
$\tau$, the ``lifetime'' of the dimer; the results are summarized in Table~\ref{tab:bound}.\\
\\
Our calculations show that all of the bound states have near-zero $P_{LZ}$ values and thus 
near-infinite lifetimes. Although there is strong coupling between the well and the states
below it, the energy gap between the well and the lower curves is too large for dissociation
to occur. In addition, the oscillation speeds of the dimers are too slow for diabatic
transitions. We therefore conclude that these dimers are stable with respect to predissociation and
so their lifetime is limited only by the Ryberg atoms themselves
($t_{\rm Ryd}\sim 700$ $\mu$s for $n=70$)~\cite{RydLife}.
\section{Conclusions}
\label{sec:conc}
In this paper, we investigated the long-range interactions between rubidium and potassium, where
both atoms are excited to high-$n$ Rydberg states. We explored all possible $\Omega$-symmetries
for both atoms being excited to the $70s$ state and the $70p$ state. Our calculations showed that
due to the effects of electronic $\ell$-mixing,
the potential energy curves describing these interactions are intricate and complicated,
particularly when the nuclear separations are in the 40,000 $a_0$ - 60,000 $a_0$ range. 
In addition, when both atoms are excited to either the $70s$ state or the $70p$ state, 
both the $\Omega=0^+$ symmetry interactions and the $\Omega=0^-$ symmetry interactions 
result in potential wells capable of supporting many bound states. 
We analyzed these wells in detail, calculating the bound
vibrational levels, the stability of these levels, and various $n$-scaling relations.

Given the interest in photoassociation experiments between different alkali species, these
results could be useful to further ultracold experiments, quantum chemistry calculations,
and/or quantum information research.
Furthermore, it might be possible to exploit the 
$\ell$-mixing for application to ``dressed'' Rydberg states~\cite{RydDress2010,RydDress2016}.
%

\end{document}